\documentclass[10pt]{amsart}
\pdfoutput=1
\usepackage{graphicx,amssymb,amsfonts,amsmath,amsthm,newlfont}
\usepackage{epsfig,url,endnotes}
\usepackage{color}
\usepackage{enumitem}
\usepackage{graphics}

\usepackage{axodraw4j}

\usepackage[all,2cell]{xy} \UseAllTwocells \SilentMatrices

\newtheorem{thm}{Theorem}[section]

\theoremstyle{definition}

\newtheorem{conj}{Conjecture}

\theoremstyle{remark}

\numberwithin{equation}{section}

\newcommand{\Z}{\mathbb Z}
\newcommand{\C}{\mathbb C}

\newcommand{\R}{\mathbb R}

\newcommand{\zetam}{\zeta^{ \mathfrak{m}}}

\newcommand{\Q}{\mathbb Q}
\newcommand{\Li}{\mathrm{Li}}

\newcommand{\To}{\longrightarrow}

\newcommand{\A}{\mathbb{A}}

\newcommand{\Gg}{\mathcal{G}}

\newcommand{\mm}{\mathfrak{m} }

\newcommand{\HF}{\mathcal{FP} }

\begin{document}
\author{Francis Brown}
\begin{title}[Periods and  Feynman amplitudes]{Periods and Feynman amplitudes}\end{title}
\maketitle

\begin{abstract}Feynman amplitudes in perturbation theory form the basis for most predictions in particle collider experiments. The mathematical quantities which occur as amplitudes include values of the Riemann zeta function and relate to fundamental objects in number theory and algebraic geometry.  This talk reviews some of the recent developments in this field, and explains how new ideas from algebraic geometry have led to much progress in our understanding of  amplitudes. In particular, the idea that certain transcendental numbers, such as $\pi$, can be viewed as a representation of a group, provides a powerful framework to study amplitudes which reveals many hidden structures. 
\end{abstract}
 \vspace{0.1in}
 
 These notes  are a faithful reproduction of the plenary talk  delivered at the international congress of mathematical physics in Santiago de Chile  in July 2015.  I have resisted the temptation to expand on the material presented in the lecture. Instead,
 I have added further details, historical notes and technical comments  in the form of optional notes at the end of the text.  Some of the material upon which this talk is based, notably on motivic periods and the cosmic Galois group, were  presented during a lecture series \cite{IHESyoutube, Cosmic, NotesMot} at the IHES in May 2015.

 It is a pleasure to thank the organisers of the ICMP, especially Rafael Benguria,  for a very successful and memorable conference.

 \section{Introduction}
 
 \subsection{Motivation}
 A fundamental  problem in high-energy physics is to understand the scattering of particles. 
 Suppose that we have a particle accelerator  in which two particles, depicted by the arrows  below on the left, are brought together at  high energy. 
 They will interact in some  way, and   two  new particles, shown leaving  on the right, are subsequently 
 observed in the detector.   Perturbative quantum field theory tells us that we cannot   directly determine  what has happened in between, and that  every possibility must be taken into account.  \vspace{0.1in} 

 \begin{center}
 {\includegraphics[width=0.4\textwidth]{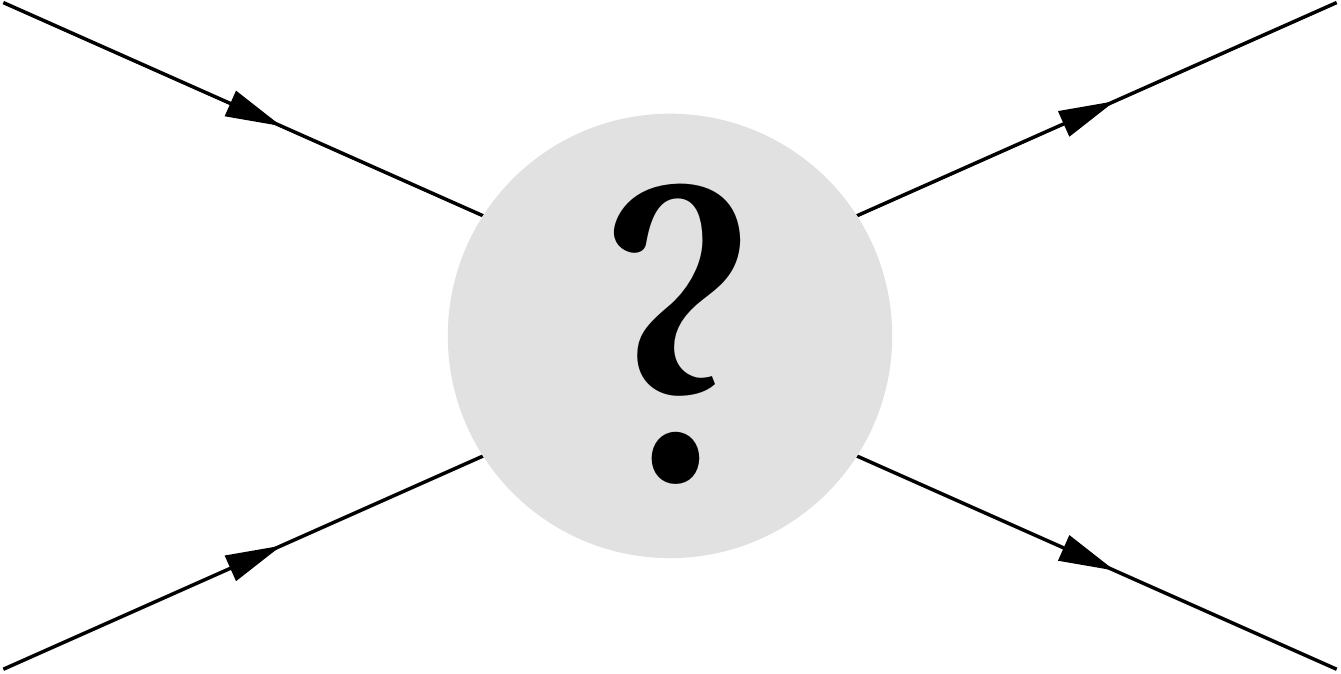}}
 \end{center}
 
The figures below depict two such possible particle interactions. In the first,  the two particles exchange a photon; in the second,  
the exchanged photon momentarily becomes an electron-positron pair which subsequently annihilate to form a photon  again.
 None of these intermediary particles can be observed.
  \begin{center}
 {\includegraphics[width=0.4\textwidth]{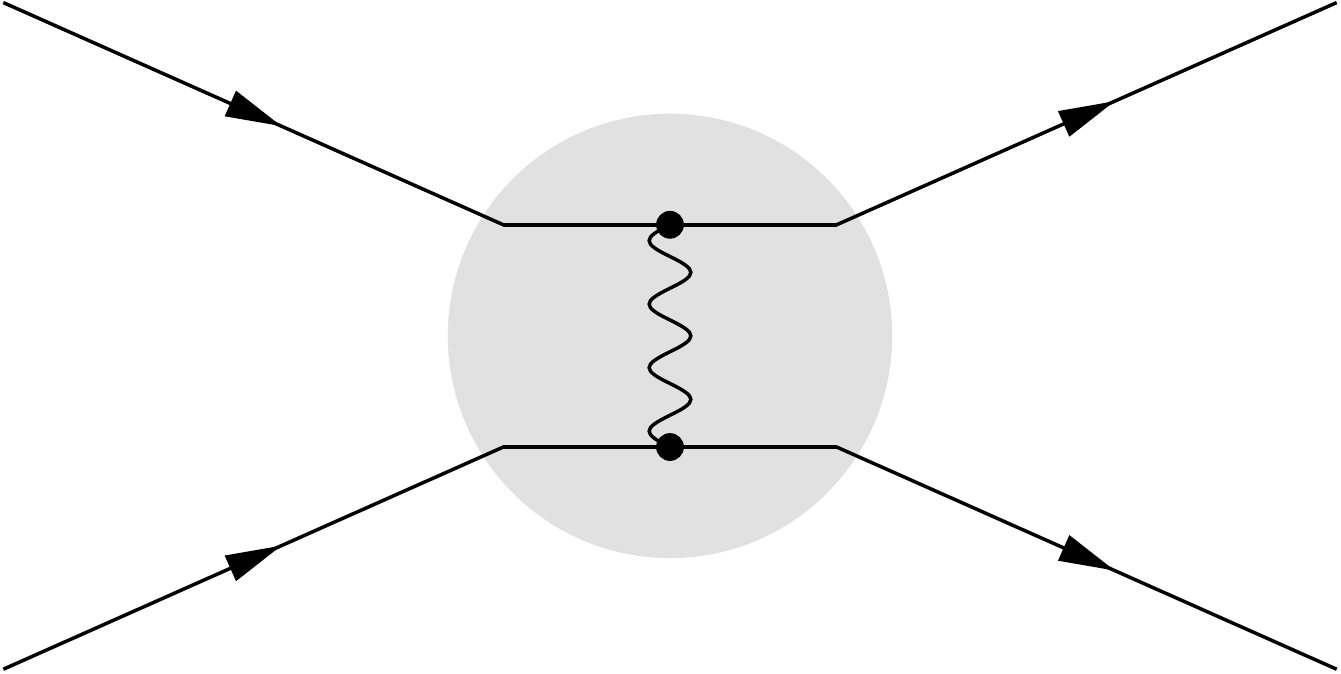}} \qquad  \qquad {\includegraphics[width=0.4\textwidth]{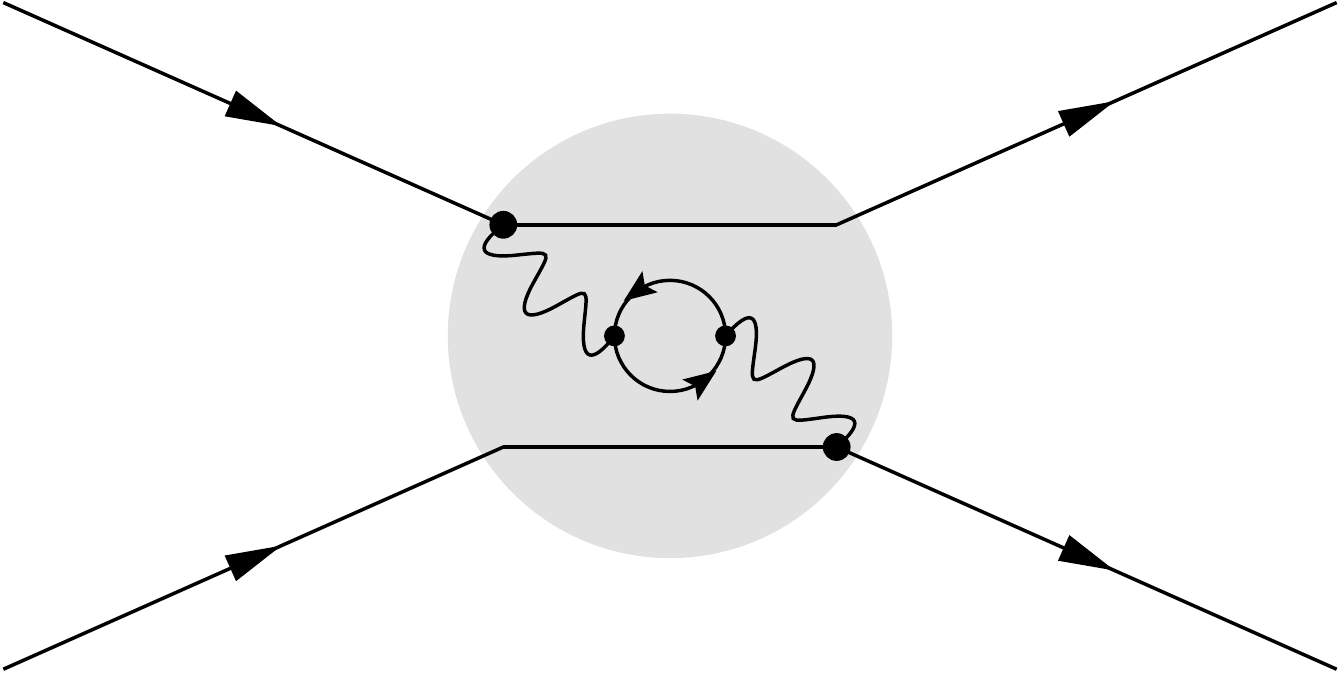}}
 \end{center}
 
 In this manner, every possible particle interaction is encoded by a graph called a Feynman diagram, to which one assigns  a probability amplitude called the Feynman amplitude.
 The  prediction for observing a particular outcome   is obtained by  adding  together  the amplitudes of all possible graphs. In practice, of course, one can only consider a finite number of graphs, and  it is expected  that the truncated sum   gives an  approximation to the true amplitude\endnotemark[1].

A first application is to provide  predictions  for particle accelerators such as the Large Hadron Collider.
The following figure  depicts some data from the LHC relating to the discovery of the Higgs boson 
taken from an online  
newspaper  
in 2013 \cite{Guardian}.
The experimental data are given by the black dots, 
 which clearly form a bump in the middle
compared  to the (blue) curve, which is called the background. The background is the theoretical prediction of the standard
 model without a Higgs boson.  The bump therefore represents the discovery of a new particle.
 
 \vspace{0.1in}
\begin{center}
{\includegraphics[width=1\textwidth]{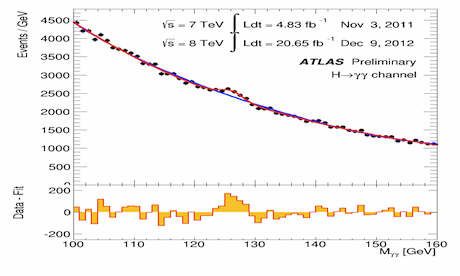}}%
\end{center}

\noindent 
The background  is obtained by calculating a huge number of Feynman amplitudes. This  is an enormous task 
 involving a large community of physicists. Although this theoretical aspect of particle discovery  is extremely  important, it does not quite receive the same  media attention  as the experimental side.

 A rather different application of Feynman diagrams relates to the anomalous magnetic moment of the electron.
 This is one of the most accurately measurable physical quantities  in nature.
 The basic Feynman diagram is depicted below on the left, and represents a slow-moving electron which
 emits a photon.  

\begin{center}
\qquad {\!\!\!\!\!\!\!\!\!\!{\includegraphics[width=0.2\textwidth]{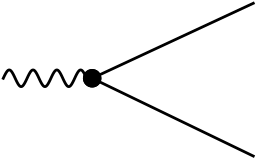}}} 
\qquad  \qquad {{\includegraphics[width=0.2\textwidth]{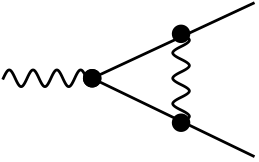}}} \qquad \qquad  %
{\put(-100,20){$+$} \put(-15,20){$ \qquad + \quad \cdots$}}
\end{center}

\noindent
The  (one-loop\endnotemark[2]) diagram on the right depicts this electron emitting and reabsorbing
a second (virtual) photon, which is not detected. The most recent experimental value  for this quantity is known to an astonishing degree 
of accuracy \cite{gminus2exp}:
 $${g-2\over 2}  = 1.001 159 652 180 91 (\pm 26) \qquad\qquad \hbox{(experiment)}$$
 The prediction from quantum field theory   is presently known to be
$${g-2 \over 2} = 1.001 159 652 181 13 (\pm 86) \qquad\qquad \hbox{(theoretical)} $$
It is  remarkable that this prediction, obtained by calculating Feynman amplitudes (as it were, by  pure thought)  currently agrees with the experimental value  to twelve digits.
 This is arguably one of the  triumphs of quantum field theory.\endnotemark[3] 

The one-loop Feynman amplitude was computed by J. Schwinger in 1947.
The seven two-loop diagrams are depicted here 
\begin{center}
{\includegraphics[height=1.5cm]{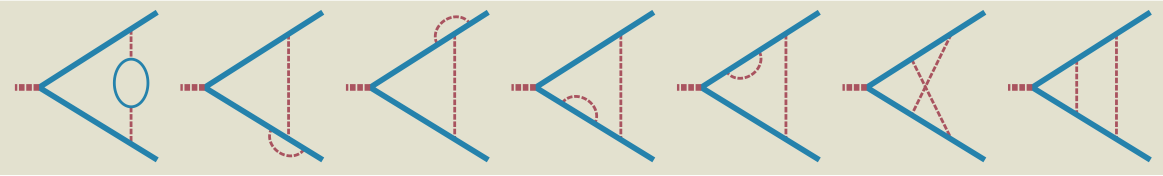}} 
\end{center}
and were correctly  computed for the first time  a decade later in 1957.
The sum of the Feynman amplitudes\endnotemark[4] for these graphs is
$$
 {197 \over 144}  + {1\over 2} \zeta(2 )   - 3\,  \zeta(2) \log 2 + {3 \over 4}  \zeta(3) \ .
 $$
 It involves the logarithm of 2, and, unexpectedly, values of the Riemann zeta function, about which we shall say a great deal more below.
 There are 72 diagrams with three loops, and the answer is   known analytically as from 1996 \cite{Laporta-Remiddi}. 
The four-loop amplitude is presently only known \emph{numerically} and involves 891 diagrams.
 Since 2012  the five-loop QED contributions\endnotemark[5] are known numerically using supercomputers due to a heroic effort 
 of T. Kinoshita and co-workers. The number of diagrams involved is of the order of  twelve thousand  \cite{Kinoshita}.
This example illustrates   two  of the main problems of perturbative quantum field theory: first of all, the sheer difficulty of computing 
Feynman amplitudes at higher loops, and secondly, the rapid proliferation of Feynman graphs as the number of loops increases\endnotemark[6].

\subsection{Zeta values}
It is a general and quite astonishing fact that amplitudes often involve 
 values of the  zeta function
\begin{equation} \label{eqn: zetadefn}
\zeta(n) = \sum_{k\geq 1 } {1 \over k^n}\qquad , \qquad  n \geq 2 \ .
\end{equation} 
The Basel problem, famously posed by Mengoli in 1644, asked for a formula for the quantity $\zeta(2)$. 
This was solved by   Euler in the 1740's  who went further and showed that  the even zeta values are all rational multiples of powers of $\pi$:
$$\zeta(2) = {\pi^2 \over 6}\quad \ ,\quad   \  \zeta(4) = {\pi^4 \over 90}\quad  \ , \ \quad  \zeta(6) = {\pi^6 \over 945}\quad \ , \ldots$$
Indeed he gave a closed formula\endnotemark[7] for every even zeta value in terms of $\pi$.
However, no such formula is known for the odd zeta values  and it is conjectured that the  $\zeta(2n+1)$, for $n\geq 1$, should be algebraically independent of $\pi$ (but it is not even known at present whether    $\zeta(3)\pi^{-3}$ is irrational).

 Although it is impossible to know what his intentions were, it seems plausible that Euler   attempted to find polynomial
relations between zeta values. In order to do this, one must  multiply together terms such as $(\ref{eqn: zetadefn})$. This quickly leads one to 
consider   double  sums which can in turn be expanded into  linear combinations of  sums with constraints on the region of summation.  In this manner\endnotemark[8],
one is quickly led to what are now called \emph{multiple zeta values}  (MZV's), defined by nested sums
$$\zeta(n_1,\ldots, n_r) =\sum_{0< k_1 < \ldots < k_r} {1 \over k_1^{n_1} \ldots k_r^{n_r}}  \qquad , \qquad n_r \geq 2\ ,$$
which depend on several indices. Curiously, these numbers were promptly  forgotten\endnotemark[9] and were ignored for over two centuries before resurfacing simultaneously in physics and several branches of mathematics in the 1990's. 

\section{Amplitudes in parametric form}
In order to simplify matters, let us replace wavy lines with straight lines in the one-loop Feynman diagram contributing  to the anomalous magnetic moment of the electron depicted above. 
It becomes the following graph.
\begin{center}
{{\includegraphics[width=0.3\textwidth]{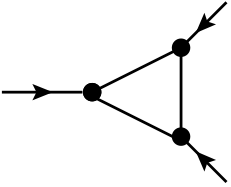}}}%
{\put(-100,34){{\small $q_1$}}     
\put(2,70){{\small $q_2$}} \put(2,2){\small $q_3$} }
{   \put(-15,34){{\small $1$}}  \put(-40,18){{\small $2$}}\put(-40,49){{\small $3$}} }
\end{center}

Thus a (scalar) Feynman diagram is simply a graph with a certain number of external half-edges, also known as external legs.
The external legs represent incoming or outgoing particles, and are assigned a momentum
$$q_i \in \R^4$$
which is a vector in  Euclidean\endnotemark[10] four-space. All particles will be represented by arrows pointing inwards: an outgoing particle
is simply an ingoing particle with negative momentum. These  momenta are subject to \emph{momentum conservation}
\begin{equation} \label{MC} 
\sum_{i } q_i = 0 \ .
\end{equation} 
To such a graph $G$ we wish to assign an integral. It is built out of certain polynomials, called Symanzik polynomials, in  variables 
$\alpha_e$ associated to  every internal edge  $e$ of the graph $G$ (known as Schwinger parameters).
The first Symanzik polynomial does not depend on the external momenta and is defined by 
\begin{equation} \label{eqn: PsiGdefn} 
\Psi_G = \sum_{T \subset G} \prod_{e \notin E(T)} \alpha_e
\end{equation} 
where the sum is over all spanning trees of $T$.  A spanning tree is a connected acyclic subgraph of $G$ which  contains  the set
of all vertices of $G$. The product is over
all edges not contained in the set of edges $E(T)$ of  $T$.   This polynomial was first introduced by Kirchhoff in order to study resistance in 
electrical circuits \cite{Kir}. 

The second Symanzik polynomial is a function of the external momenta (denoted simply by $q$ for brevity) and is defined by 
$$\Phi_G(q) = \sum_{T_1 \cup T_2 \subset G}     (q^{T_1})^2  \prod_{e \notin E(T_1) \cup E(T_2)} \alpha_e $$
where the sum is over all spanning 2-trees: that is to say, subgraphs with exactly two components $T_1$, $T_2$, each of which
are trees, and such that every vertex of $G$ is contained in exactly one such tree. 
For a four-vector $q= (q_1,q_2,q_3,q_4) \in \R^4$ its square is respect to the Euclidean norm
$$q^2 = q_1^2 + q_2^2 +q_3^2 +q_4^2$$
and $q^{T_1}$ denotes the total momentum entering the component $T_1$.  By momentum conservation 
$q^{T_1} $ is equal to $ -q^{T_2}$ and  therefore  $(q^{T_1})^2=(q^{T_2})^2$.

The particle masses enter the picture via a polynomial we shall denote by 
\begin{equation} \label{eqn: Xidefn} 
\Xi_G(q,m) = \Phi_G(q) + \big(\sum_{e\in E(G)} m_e^2 \alpha_e\big)\,  \Psi_G\ .
\end{equation}
Here $m_e$ is a particle mass assigned to each internal edge of $G$.  

For example, consider the graph above. It has three spanning trees
\begin{center}
{\includegraphics[width=0.25\textwidth]{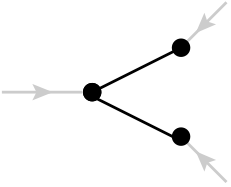}} \qquad 
{\includegraphics[width=0.25\textwidth]{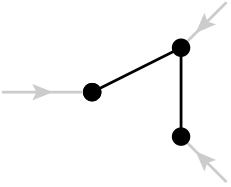}} \qquad
{\includegraphics[width=0.25\textwidth]{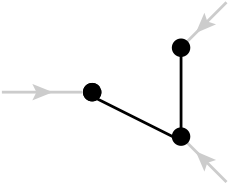}}%
{\put(-15,34){{\small $1$}}}{\put(-40,18){{\small $2$}}}{\put(-40,49){{\small $3$}}}
{\put(-130,34){{\small $1$}}}{\put(-155,18){{\small $2$}}}{\put(-155,49){{\small $3$}}}
{\put(-243,34){{\small $1$}}}{\put(-267,18){{\small $2$}}}{\put(-267,49){{\small $3$}}}
\end{center}
and so $\Psi_G = \alpha_1+\alpha_2+\alpha_3$.  It has the following spanning 2-trees:
\vspace{0.1in}
\begin{center}
{\includegraphics[width=0.25\textwidth]{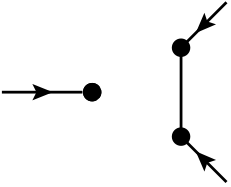}} \qquad 
{\includegraphics[width=0.25\textwidth]{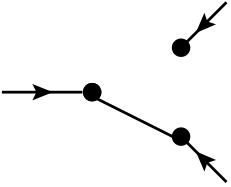}} \qquad
{\includegraphics[width=0.25\textwidth]{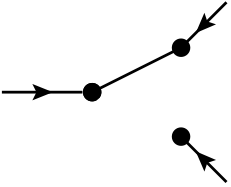}}%
{\put(-15,34){{\small $1$}}}{\put(-40,18){{\small $2$}}}{\put(-40,49){{\small $3$}}}
{\put(-130,34){{\small $1$}}}{\put(-155,18){{\small $2$}}}{\put(-155,49){{\small $3$}}}
{\put(-243,34){{\small $1$}}}{\put(-267,18){{\small $2$}}}{\put(-267,49){{\small $3$}}}
{\put(-305,27){{\small $q_1$}}}{\put(-122,56){{\small $q_2$}}}{\put(-15,0){{\small $q_3$}}}
\end{center}
which gives $\Phi_G(q) =q_1^2 \alpha_2\alpha_3 + q_2^2 \alpha_1 \alpha_3 + q_3^2 \alpha_1\alpha_2 $.  
Therefore in this case, 
\begin{multline} \Xi_G(q,m) = q_1^2 \alpha_2\alpha_3 + q_2^2 \alpha_1 \alpha_3 + q_3^2 \alpha_1\alpha_2 +    \\
 ( m_1^2 \alpha_1 + m_2^2 \alpha_2 + m_3^2 \alpha_2) (\alpha_1 + \alpha_2 + \alpha_3) \ .
\end{multline} 
Let $h_G$ be the number of loops,  and  $N_G$  the number of edges of $G$.
In general,  $\Psi_G$ and $\Phi_G(q)$ are homogenous in the $\alpha_e$ of degree equal to $h_G$ (respectively $h_G+1$),
and therefore $\Xi_G(q,m)$ is also homogeneous in the Schwinger parameters. 

We can now write down the amplitude\endnotemark[11]  in parametric form as the integral
$$ I_G(q,m) = (*) \int_{0< \alpha_e <\infty}  {1 \over \Psi_G^{2} } \Big( {\Psi_G \over \Xi_G(q,m)} \Big)^{N_G-2h_G} \Omega_G $$
where  $(*)$ are certain  Gamma-factors which we omit, and
$$\Omega_G = \sum_{e} (-1)^e \alpha_e d\alpha_1 \ldots \widehat{d \alpha_e} \ldots d\alpha_{N_G}\ .$$
We shall simply take  $(*)$ to be equal to $1$. The notation is slightly inaccurate:
in the physics literature the integral is often written with a Dirac delta function inside the integrand; mathematicians prefer
to think of it  as an integral in projective space over the real coordinate simplex of dimension equal to $N_G-1$ (one checks that the integrand is indeed homogeneous of degree $0$).  There are two important points to bear in mind. The first is that the integrals often diverge.
The theory of \emph{renormalisation} provides a way to remove ultra-violet singularities  in a consistent way. Infra-red singularities pose more subtle
 problems. The second observation is that  amplitudes in any gauge theory (such as quantum electrodynamics)  can always be written in parametric form. This will produce  polynomials in the  numerator, but will not change the form of the denominators (see e.g., \cite{Corolla}).

\subsection{Generalized Feynman amplitudes} In order to include the possibility of gauge theories, 
we can consider the following large class of integrals, which we call  `generalised amplitudes'  (this terminology is not standard), defined by any integral
\begin{equation} \label{eqn: GenAmp}
\int_{0< \alpha_e <\infty}  {P (\alpha_e) \over (\Psi_G)^A (\Xi_G(q,m))^B}\,  \Omega_G 
\end{equation}
where  $P(\alpha_e)$ is a homogeneous polynomial in the $\alpha_e$,  and $A$ and $B$ are integers,
such that the integrand is homogeneous of degree $0$ and converges. 
The key point, as we shall see below, is that  the  integral can be defined entirely in terms of \emph{polynomials}. It is a  fact, which is not obvious\endnotemark[12], that the type of quantities  which can occur as such integrals are determined by the  denominator,  and not the numerator $P$.

As implied by the previous discussion, amplitudes are extremely  difficult to compute. 
The amplitude of the triangle graph (above) can be expressed using the dilogarithm function, first defined by  Leibniz:
$$\Li_2(x) = \sum_{n\geq 1 } {x^n\over n^2}\ .$$
Its  arguments are square roots of algebraic functions of  the masses $m_e^2$ and external momenta $q_i^2$ \cite{Nickel, Davidychev, BlochKreimer}.   However, not all two-loop amplitudes are  known functions in the sense that they probably do not  belong to the canon of classical special functions,
and  their general analytic properties are not completely understood.

\subsection{Some examples which are numbers}
We have already seen the example of the anomalous magnetic moment of the electron, in which
the amplitudes are essentially numbers and depend in a trivial way on masses and momenta. Such a process could be called a no-scale (or single-scale) process. 
Another such situation  occurs when $N_G=2h_G$, and we can consider the projective integral
\begin{equation}  \label{eqn: phi4int}
 I_G = \int_{0< \alpha_e<\infty} {\Omega_G \over \Psi_G^2} 
 \end{equation} 

\noindent
which does not involve  the polynomial $\Xi_G$. One can show that it converges if $G$ is  \emph{primitive} ($N_{\gamma} > 2h_{\gamma}$ for all 
strict subgraphs $\gamma \subsetneq G$). The numbers $(\ref{eqn: phi4int})$  are of physical relevance in that they are  renormalisation-scheme independent contributions to the 
 beta function in massless $\phi^4$ theory in four space-time dimensions.  The condition of being in $\phi^4$ theory is equivalent to the statement
 that the valency of each vertex is at most four. The amplitudes in this theory are known to much higher loop orders \cite{BK, Census} than most other 
 quantum field theories and provide a useful testing ground for conjectures. 

Below are examples of primitive graphs in $\phi^4$ at 3,4,5 and 6 loops which have convergent integrals, together with the values $(\ref{eqn: phi4int})$  immediately underneath. 
\begin{center}
\quad {{\includegraphics[width=7cm]{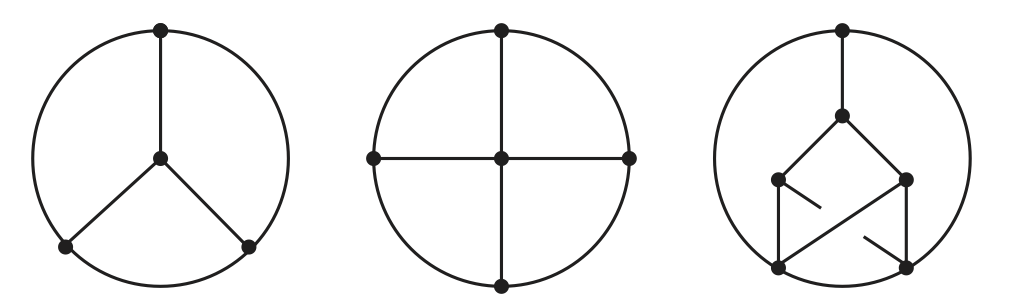}} 
{\includegraphics[width=1.8cm]{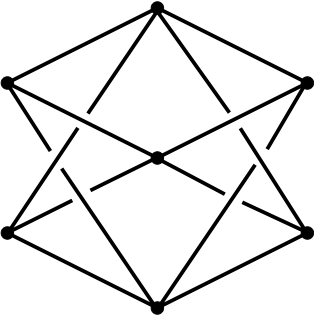}}}%
\end{center}

$$I_G: \!\!\! \quad 6 \zeta(3)  \qquad \qquad 20 \zeta(5)  \qquad \qquad 36 \zeta(3)^2 \qquad  \quad 32 \, P_{3,5}$$
Here the quantity  $P_{3,5} =\textstyle{- {216\over 5} \zeta(3,5)  -81 \zeta(5)\zeta(3) +{522\over 5} \zeta(8)}$ and involves multiple zeta values.
This fact was  observed experimentally in \cite{BK}.

\section{Recent results}

 It is impossible, during the space of a single talk, to do to justice to the spectacular recent progress in our understanding of amplitudes. 
According to  received wisdom,    amplitudes in quantum field theories should correspond 
to simple mathematical objects. However, recent results in massless $\phi^4$ theory apparently  contradict this belief and 
tell a cautionary tale  which 
probably reflects a general phenomenon which also holds for   other quantum field theories.

In this particular situation (massless $\phi^4$),  it was believed for a long time, based on a vast amount of evidence,   that all primitive amplitudes of the form  $(\ref{eqn: phi4int})$ should be linear
combinations of  multiple zeta values, which, as we have seen, were  introduced  several hundred years earlier   and for entirely different reasons. The hope was that  certain fundamental quantities arising in mathematics and physics should be one and the same. 

However, after a considerable amount of hard work over the last decade \cite{Sta, Stem, BB, BEK, BrFeyn, DorynPoints, K3phi4, BrownDoryn}, we now know that this is false: there are explicit examples of amplitudes in $\phi^4$ at high loop orders (and quite beyond the  reach of  our present computational ability), which are related to 
 modular forms. These are not  expected to be expressible as multiple zeta values. 
   Furthermore, not all multiple zeta values actually appear as amplitudes (there are  `gaps', as we shall see later).   The numbers $\zeta(2)$ or $\zeta(4)$, for example, cannot be obtained as integrals of the form $(\ref{eqn: phi4int})$.

Similarly, the fact that the general two-loop amplitude (with arbitrary numbers of masses and external momenta) has also 
   resisted classification is further evidence of  the complicated nature   of  amplitudes.

In brief, one cannot hope for a simple answer to describe amplitudes at all loop orders, and they are mathematically  considerably  richer  than  previously imagined. 

\subsection{Recent methods}
Despite this apparent pessimism, there has been a proliferation of new techniques\endnotemark[13] in the study of amplitudes over the last few years.
These have lead to  impressive progress in the theoretical understanding of amplitudes.  The main  broad themes are:

\begin{enumerate}
\item New  insights  from algebraic geometry (for example,  $p$-adic methods, Hodge theory, and the theory of motives), and number theory (e.g., the modularity programme which has inspired many guesses for amplitudes).
\item Techniques from the theory of iterated integrals. Examples include  the use of  symbols,  and  unipotent differential equations. The systematic use of  multiple and,  more recently, elliptic polylogarithms (which are special cases of the theory of iterated integrals) have led to organising principles for certain classes of amplitudes with many  practical applications. 

\item  New geometric interpretations of  amplitudes.  New interpretations of scattering amplitudes in terms of    Grassmannian geometry, cluster algebras, twistors, moduli spaces of curves, or graph hypersurfaces  have each led to striking  results, and new ways to think about amplitudes. 
\end{enumerate}
The diversity of techniques listed above illustrates how the study of amplitudes is a very active and rapidly expanding field.
However, every one  of the  approaches listed above runs into the same difficulty\footnote{
 A possible exception is the planar limit of $N=4$ SYM where  an all-orders expression for the amplitude
 may  shortly be within reach.}, namely that  each new loop order involves mathematical objects which are an order of magnitude more complex than the last. Indeed, many entirely reasonable conjectures have turned out to be false after examining examples with  just one  more loop.  So although  our understanding of amplitudes at low loop orders has been hugely simplified using modern techniques, 
the  unavoidable fact  is that arbitrary  amplitudes remain as out of reach as ever. 
\vspace{0.2in}

\subsection{Where do we go from here?} Faced with these difficulties,  what could we possibly say that is true for all quantum field theories, and at all loop orders? Are there \emph{any} organising principles which are valid for all Feynman amplitudes?

In the second half of this talk, I want to propose a tentative answer to these  questions and present some evidence for it.
This  answer will take the following form: there should be a very large group of deeply hidden symmetries which acts on the 
space of all  generalised Feynman amplitudes in a rather subtle way.  This group will propagate information between amplitudes
of different loop orders.  But before we can start  to describe this group, we need to change radically  the way
we think about numbers and integrals. That brings us to the next topic.

\section{Periods and Galois theory}
\subsection{ Periods  and families of periods}
A \emph{period}  is a complex number whose  real and imaginary parts are integrals of  rational  differential forms 
 over  domains defined by polynomial inequalities (all with rational coefficients)
$$ I = \int_{\sigma} {P \over Q}\,  dx_1 \ldots dx_n \ , \qquad P, Q \in \Q[x_1,\ldots, x_n]\ . $$
This  elementary definition of a period is due to Kontsevich and Zagier \cite{KoZa}. For the study of Feynman amplitudes, we need
some greater flexibility and we call 
 $I$ a  \emph{family of periods} when  $P,Q$ or $\sigma$ depend algebraically on parameters. One way this can happen  is if  $P$ or $Q$ depend on variables which 
 are not integrated out. Examples of periods, and families of periods include:
\begin{itemize}
\item  All algebraic numbers, such as 
$$ \sqrt{2} = \int_{x^2 \leq 2} {dx \over 2}\ .$$
\item Many classical transcendental numbers such as
$$\zeta(2) =      \int_{0\leq x,y \leq 1} {dx dy \over 1-xy}  \ . $$
Indeed one easily shows that all zeta values and multiple zeta values\endnotemark[14] are periods according to this definition.
\item Many classical functions such as
$$ \log y  =  \int_{1\leq x\leq y} {dx \over x} \ .$$
Here the variable $y$  occurs in the polynomial inequalities defining the domain of integration.
\item Convergent  generalised Feynman amplitudes $(\ref{eqn: GenAmp})$.
\end{itemize} 
One can show that  families of periods satisfy  Picard-Fuchs  differential equations.

\subsection{Towards a Galois theory for periods}
 E. Galois in around 1830 introduced ideas which are now collectively known as Galois theory. In its modern formulation, the fundamental object is a group\endnotemark[15] $\mathrm{Gal}(\overline{\Q}/\Q)$ which is the group of symmetries of the ring of algebraic numbers which respects the operations of addition and multiplication.  It therefore permutes the roots of any polynomial equation.

  For example, the number $\sqrt{2}$ is 
 defined as a solution to 
 $$X^2- 2= 0\ .$$
  But this equation has another solution, namely $-\sqrt{2}$, which is indistinguishable from $\sqrt{2}$ from the equation alone. 
  Galois' idea is that the inherent  ambiguity in the definition `the square root of 2' is accounted for  by the existence
  of a group of symmetries. In particular, every element $\sigma$ in
$\mathrm{Gal}(\overline{\Q}/\Q)$ permutes  $\sqrt{2}$ and    $- \sqrt{2}$.

Given that periods contain all algebraic numbers, could it be the case that Galois theory generalizes to all periods? Y. Andr\'e has suggested exactly that \cite{An1, An2}.  He observed that 
the  existence of the conjectural category of mixed motives together with  a  version of  Grothendieck's period conjecture would imply 
the existence of such a theory. Furthermore, he showed that several classical conjectures in transcendental number theory would
follow as  consequence of such a theory \cite{An3, Bert}.

A fundamental  problem with this programme is that it is impossibly conjectural.\endnotemark[16] Grothendieck's period conjecture is one of the most ambitious and difficult open problems 
in mathematics: proving  the transcendence of $\zeta(3) \pi^{-3}$, for example, would be a very  special case. 
However,  I claim that we can circumvent all conjectures in certain specific situations using  a   notion called motivic periods, and
that this works in the case of Feynman amplitudes.

A motivic period is an object which naturally carries the action of a certain group.  The definition that I will use in fact has nothing to do with 
the theory of motives. It is  motivic only in the sense that, were all the above conjectures to be true, then the group action would exactly calculate the correct one predicted by the philosophy of motives. 
What follows is therefore a 
slight shift in perspective:  we will replace periods or families of periods with marginally different objects which capture more of their structure. This   is best illustrated on a simple example.

\subsection{The motivic period corresponding to $2 \pi i $.}
Recall that  Cauchy's theorem is equivalent to the statement\endnotemark[17] 
$$2 \pi i = \int_{\gamma} {dx \over x}\ ,$$
where $\gamma$ is a small loop in  $\C^{\times}$ which winds positively  around the origin. 
The integral is to be interpreted as follows. The integrand is a closed one-form on $\C^{\times}$. By Stokes' theorem, the integral
is unchanged if we add to it  an exact differential, so we should consider its class modulo exact one-forms. The vector space of closed forms modulo exact forms  defines ordinary de Rham cohomology $H^1(\C^{\times})$. It turns out, in this particular case\endnotemark[18], that Grothendieck's \emph{algebraic} de Rham cohomology  is given by the naive analogue where one considers algebraically-defined forms:
$$\Big[{dx \over x}\Big] \quad \in \quad  {\hbox{Closed \emph{algebraic} 1-forms}  \over \hbox{Exact  \emph{algebraic} 1-forms}}  = H^1_{dR} \cong  \Q[\textstyle{dx \over x}]$$
In this case it is a one-dimensional vector space over $\Q$, spanned by the class of ${dx \over x}$. 
Similarly, the domain of integration is a closed chain, and again by Stokes' theorem, can be modified by the boundary of two-chain. Hence we only care about its class in the quotient, which is the singular or Betti homology:
$$[\gamma] \quad \in \quad  {\hbox{Closed 1-chains}  \over \hbox{Boundaries of  2-chains}} = (H_B^1)^{\vee}  
\cong \Q[\gamma] $$
It is the dual to the Betti cohomology (with rational coefficients), and in this case is a one-dimensional vector space over $\Q$, spanned by the class of  $\gamma$.

Now integration is simply a way to pair integrands and domains of integration to obtain a number. 
Therefore Grothendieck's interpretation of integration \cite{Groth} is an isomorphism between two vector spaces:
\begin{eqnarray}
\mathrm{comp} : \quad H^1_{dR} \otimes_{\Q} \C   & \overset{\sim}{\To}  & H^1_B \otimes_{\Q} \C \nonumber \\ 
\omega & \mapsto &  \textstyle{(\gamma \mapsto \int_{\gamma} \omega) } \nonumber 
\end{eqnarray} 
We are forced to tensor with the complex numbers because integration of course produces numbers which are not always rational.

We can now define the `motivic' period corresponding to $2\pi i$.
Recall that the square-root of $2$ was represented by  algebraic data: the equation $X^2-2=0$.  
The analogue of Cauchy's integral representation for $2\pi i$ is the  data
$$  (( H^1_B, H^1_{dR} ,   \mathrm{comp}) , [\gamma], [\textstyle{dx\over x}]  )  \ .$$
This is essentially what it means to write down an integral: the integrand is the differential form ${dx/x}$, which defines a class in  a vector space $H^1_{dR}$, 
the domain of integration is the chain $\gamma$, which defines a class in the dual of a vector space $H^1_B$, and the integral symbol $\int$ is encoded by the comparison map.

The \emph{motivic period}\endnotemark[19] is the equivalence class of this data modulo some equivalence relation.  It is denoted by a superscript $\mathfrak{m}$:
$$(2 \pi i )^{\mm}: =  \big[(H^1_B ,  H^1_{dR} ,   \mathrm{comp}) , [\gamma] ,  [\textstyle{dx\over x}] )  \big]\ .$$
The number $2 \pi  i$ can be  retrieved from this data by performing the integral
$$\mathrm{per} ( (2 \pi  i )^{\mm}) =  [\gamma]  \big( \mathrm{comp} \, [\textstyle{dx\over  x}]\big)  =  2 \pi i \ . $$
This is called the period map. Motivic periods form a ring which carries many extra structures.\endnotemark[20]

We can define the \emph{Galois group  $\Gg$ of motivic periods} as the group of symmetries of their defining data. It is an affine group scheme, whose rational points will be denoted $\Gg(\Q)$. Hereafter it will simply be called  the ``Galois group" for simplicity, but it is not to be confused with the classical Galois group of algebraic numbers $\mathrm{Gal}(\overline{\Q}/\Q)$; there is a surjective homomorphism $\Gg(\Q)\rightarrow \mathrm{Gal}(\overline{\Q}/\Q)$.  This uses
some basic theory of Tannakian categories.  In this case, $\Gg(\Q)$  acts via 
$$ (2 \pi i)^{\mm} \mapsto \lambda  (2 \pi i)^{\mm} \qquad \hbox{ for some } \lambda \in \Q^{\times}\ .$$

\subsection{Further examples}
The action of the group $\Gg(\Q)$   on the motivic period   $(2  \pi i )^{\mm} $ via $(2 \pi i)^{\mm} \overset{}{\mapsto} \lambda_g (2 \pi i)^{\mm}$ defines  a one-dimensional representation of $\Gg(\Q)$:
$$ g \mapsto \lambda_g : \Gg(\Q) \To \Q^{\times}\ . $$
Similarly, it is not difficult to check that  the Galois action on the motivic period corresponding to  $\log(2)$  is  given by\endnotemark[21]  
$$ \log^{\mm}(2) \overset{}{\mapsto} \lambda_g \log^{\mm}(2) + \nu_g  .1 $$
where $\nu_g\in \Q$.
 Thus $\log^{\mm}(2)$ generates a two-dimensional representation of $\Gg(\Q)$:
$$ g \mapsto 
\left(
\begin{array}{cc}
  \lambda_g    &   0\\
  \nu_g  &    1 
\end{array}
\right) \quad : \quad \Gg(\Q) \To  \mathrm{GL}_2(\Q) \ .
$$
The dimension of the representation generated by a motivic period is  a new  invariant which  is one possible measure of   its complexity.  There is a whole raft
of new invariants  that one can define in a similar way.

Multiple zeta values have motivic versions $\zetam(n_1,\ldots, n_r)$ whose  Galois group is known \cite{BrMTZ}.   This  example  is one of  the simplest
possible  prototypes for a general Galois theory of periods\endnotemark[22], and in this case,  we know exactly which representations  can arise in this way.
For example,  the Galois group $\Gg(\Q)$ acts quite differently on odd and even\endnotemark[23] zeta values, as one might expect from  Euler's theorem: \begin{eqnarray} 
 \zetam(2n) & \mapsto & \lambda^{2n}_g\, \zetam(2n)  \nonumber  \\
 \zetam(2n+1)  & \mapsto  &\lambda^{2n+1}_g\, \zetam (2n+1)+ \sigma^{(2n+1)}_g. 1 \nonumber  
\end{eqnarray}
Here $\lambda_g$ is the same character as above, but  $\sigma_g^{(2n+1)} \in \Q$  provides  a new degree of freedom for each $n\geq 1$.
Note that  these examples  predict the \emph{conjectural} motivic Galois action on the actual numbers $\log 2, \zeta(2n), \zeta(2n+1)$, and so on.

  The power of $\lambda_g$ in the previous formulae defines a grading on motivic multiple zeta values which is equal to the sum of arguments 
  $n_1+\ldots +n_r$ and is called the weight. For general motivic periods, this is not a grading, but a filtration. 
The multiple zeta value $\zetam(3,5)$  defines  a 3-dimensional representation of $\Gg$:
$$ g   \left(
\begin{array}{cccc}
  \zetam(3,5)&     
  \zetam(3)  &
1
\end{array}
\right)
 =   
 \left(
\begin{array}{cccc}
  \zetam(3,5)&     
  \zetam(3)  &
1
\end{array}
\right)
  \left(
\begin{array}{ccc}
   \lambda_g^8  & 0    & 0    \\
 - 5 \lambda_g^3 \sigma^{(5)}_g &    \lambda_g^3 & 0    \\
  \sigma^{(3,5)}_g & \sigma^{(3)}_g   &   1 
\end{array}
\right)  
$$
We can start to see from the coefficient $-5$ sneaking into the first column of this matrix that the Galois action is in general  complicated
and far from obvious.

In general, the \emph{Galois conjugates} of a motivic period are the elements in the representation it generates. In the following examples
this will coincide with the space spanned by the orbit under $\Gg(\Q)$. 
The Galois conjugates of $\zetam(3,5)$, therefore, can be any linear combination\endnotemark[24]  of $1, \zetam(3), \zetam(3,5)$.
 The number $1$, for example,  is a Galois conjugate of  $\zetam(3)$, but not of $\zetam(2)$.

\section{Motivic periods corresponding to Feynman amplitudes} 
Since generalised Feynman amplitudes $(\ref{eqn: GenAmp})$ are periods\endnotemark[25], it is natural  to ask  whether 
 the theory of motivic periods can be applied to them. This is indeed possible. 
 We shall call a Feynman graph of type $(Q,M)$ if it has at most $Q$ non-zero external momenta, and at most $M$ possible non-zero particle masses.

\begin{thm} \label{thm1} For every Feynman graph $G$, and any convergent\footnote{This theorem, as presently stated, only covers the case of convergent amplitudes with mild conditions on the external kinematics.  By cobbling together various results in the literature, it is not difficult to extend it to the case  of graphs with ultra-violet subdivergences,
and incorporate the theory of renormalization.  I am confident that there is an analogous version  covering the case of graphs with infra-red singularities,  but this 
would require a more detailed analysis.} generalised amplitude of $G$, we can {\bf canonically} define the corresponding  motivic period.
Let $\mathcal{FP}_{Q,M}^{\mm}$ be  the vector space of all motivic periods\endnotemark[26] of  Feynman graphs of type $(Q,M)$. 
It admits an action of a group $\mathcal{C}_{Q,M}$,  the \emph{cosmic Galois group}.
For every convergent Feynman amplitude $I_G$  we have a motivic version  
$$I^{\mm}_G \in \mathcal{FP}_{Q,M}^{\mm} \quad \hbox{ satisfying }\quad  \mathrm{per\,} (I^{\mm}_G) = I_G \ .$$
 \end{thm}
 The group 
$\mathcal{C}_{Q,M}$ is a pro-algebraic matrix group. In this way, every Feynman amplitude of a graph $G$  of type $(Q,M)$ defines a  finite-dimensional  representation of  $\mathcal{C}_{Q,M}$. We obtain in this way a whole swath of new invariants that one can associate to Feynman amplitudes that were not previously visible.  Here we shall mention just one, which 
is the notion of weight. This takes the form of an increasing filtration on the vector space $\mathcal{FP}_{Q,M}^{\mm}$ called the \emph{weight} filtration:\footnote{We already saw that motivic multiple zeta values are graded by their weight.  The notion of weight here is twice the MZV-weight. The weight is not a grading but a filtration  in general.}  
$$W_k \mathcal{FP}_{Q,M}^{\mm} = \langle\hbox{motivic periods of weight } \leq k\rangle \ . $$
The notion of weight comes from the theory of mixed Hodge structures \cite{De1,De2}.

\subsection{Stability theorem}
 Because there  are infinitely many  Feynman graphs of type $(Q,M)$,  we expect the  space of periods they generate to be  infinite dimensional in every weight.
 However, the following theorem   implies that the low-weight periods of  graphs with arbitrarily many loops are already
spanned by the periods of a finite number of small graphs. We call this  the  principle of small graphs.

\begin{thm} \label{thm2} For every $(Q,M)$, the space
$
W_k \mathcal{FP}_{Q,M}^{\mm} $ is finite-dimensional. 
\end{thm}

This is a powerful statement about generalized amplitudes  to all  loop orders and encodes infinitely many relations between them. 
We can combine this theorem with the action of the cosmic Galois group. Suppose we 
wish to know something about the amplitude $I_G$ of a large graph. Every Galois conjugate $g I^{\mm}_G$, $g\in \mathcal{C}_{Q,M}$ which 
is of small weight must necessarily, by the previous theorem, be a period of a small graph (which, for the sake of argument,  we already  know). This can be viewed as an equation satisfied by $I^{\mm}_G$, and hence a constraint on its period $I_G$.  In fact, even knowing something about periods of Feynman graphs with merely three edges,  surprisingly  leads to strong constraints on amplitudes \emph{to all  loop orders}.\endnotemark[28]
One can be more precise: the Galois conjugates $gI^{\mm}_G$  of a Feynman amplitude which are  of weight $\leq k$ lie in a finite dimensional vector space generated
by certain motivic periods  constructed  out   of its sub-quotient graphs  (graph minors) with  $k+1$ edges or  fewer.

\subsection{Partial factorization of graph polynomials}
The  stability theorem uses the following elementary but non-trivial property of graph polynomials. Let $\gamma \subset G$ be a subgraph, and  let $G /\gamma$ be the  graph obtained by contracting $\gamma$. Then
\begin{equation} \Psi_G = \Psi_{\gamma} \Psi_{G / \gamma} + R_{\gamma,G}
\end{equation} 
where the polynomial $R_{\gamma,G}$ is of higher degree in the edge variables corresponding to the subgraph $\gamma$ than $\Psi_{\gamma}$. 
For example, consider the following graph:
\vspace{0.1in}

{\qquad \qquad {\includegraphics[width=0.2\textwidth]{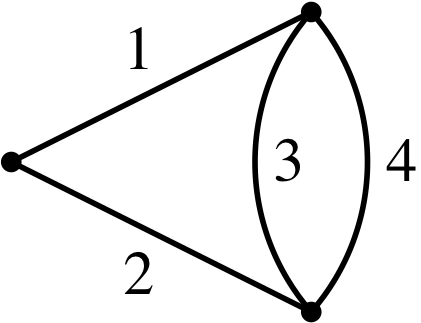}} }\qquad %
{\put(-100,26) {{$G=$}}
      {\put(50,22){{$\Psi_G = \alpha_1 \alpha_3 + \alpha_1\alpha_4 + \alpha_2 \alpha_3 + \alpha_2 \alpha_4 + \alpha_3\alpha_4$}}}}%
 
The following  choice of subgraph gives one partial factorisation:
\vspace{0.1in}

{\qquad \qquad{\includegraphics[width=0.2\textwidth]{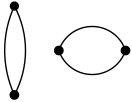}}}%
 {\put(-80,5) {{{\small $\gamma$}}}
  \put(-38,5) {{{\small $G/\gamma$}}}
   \put(-23,34) {{\tiny $1$}}  \put(-23,18) {{\tiny $2$}} 
      \put(-76,24) {{\tiny $3$}}  \put(-56,24) {{\tiny $4$}} 
      {\put(50,22){{$\Psi_G = \underbrace{(\alpha_3+ \alpha_4)}_{\Psi_{\gamma}}  (\underbrace{\alpha_1+\alpha_2}_{\Psi_{G / \gamma}} )+ \underbrace{\alpha_3\alpha_4}_{R_{\gamma,G}}$}}}} %
   
Another choice of subgraph gives a different partial factorisation:   
\vspace{0.1in}

    {\qquad \qquad {\includegraphics[width=0.2\textwidth]{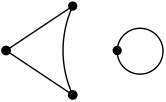}}}%
    { \put(-67,0) {{{\small $\gamma$}}}
  \put(-25,0) {{{\small $G/\gamma$}}}
   \put(-0,20) {{\tiny $4$}}  \put(-41,20) {{\tiny $3$}} 
      \put(-60,32) {{\tiny $1$}}  \put(-60,8) {{\tiny $2$}} 
      {\put(50,22){{$\Psi_G = \underbrace{(\alpha_1+\alpha_2+ \alpha_3)}_{\Psi_{\gamma}}  \underbrace{\alpha_4}_{\Psi_{G / \gamma}} + \underbrace{\alpha_1\alpha_3+\alpha_2\alpha_3}_{R_{\gamma,G}}$}}}}%
\vspace{0.1in}

This property is the crucial, and miraculous, feature of graph polynomials which allows the stability theorem to hold.
One can show that this property determines the graph polynomials $\Psi_G$ essentially uniquely. 
There are similar factorisation properties\endnotemark[29] for the graph polynomials $\Phi_G(q)$ and $\Xi_G(q,m)$.

  This factorisation property  is used  in the theory of renormalisation but only for certain subgraphs  $ \gamma$ which are called divergent.  However,  it is true for all subgraphs. This extra information is exploited in theorem $\ref{thm2}$.

\section{Applications}

\subsection{All-order results in perturbative quantum field theory}
Let  $G$  be a Feynman graph with $h_G$ loops, and  $I_G$ a convergent Feynman amplitude.
In the previous discussion, the weight  of $I^{\mm}_G$ was bounded by the \emph{number of edges} of $G$. The following conjecture
states that the weight should also be bounded as a function of the \emph{number of loops}.

\begin{conj}  \label{conj1} The weight of $I^{\mm}_G$ is less than or equal to $4h_G$. 
\end{conj}

This conjecture should follow from the momentum-space representation of the Feynman integral.
In an even number of  space-time dimensions $D$, the bound should be $Dh_G$. For the purposes of this talk, we shall take $D=4$ but the previous theorems hold when $D$ is any positive even value.

Combined with  stability (theorem $\ref{thm2}$ and remarks which follow) this implies the following statement   for graphs with many legs: 
\emph{ if $G$ has $h_G$ loops, its amplitude $I_G$ is a (possibly regularised) period of its sub-quotient graphs with $\leq 4h_G+1$ edges.}

Consider the simplest situation when $h_G=1$. In this case the conjecture is certainly true, and   the amplitudes of a one-loop graph  have weight $\leq 4$. Therefore, by theorem $\ref{thm2}$, they  are expressible in terms of the periods of one-loop graphs with $\leq 5$ edges\footnote{in fact, one can do a little better in this case, but this is the generic answer.}. Below on the left  is a graph with one loop and  eight external legs. Its sub-quotient graphs with $\leq 5$ edges (obtained by deleting and contracting edges) are pentagons, squares, and so on, down to 2 and 1 edge graphs (not shown).
\vspace{0.1in}

\begin{center}

{{\includegraphics[width=1\textwidth]{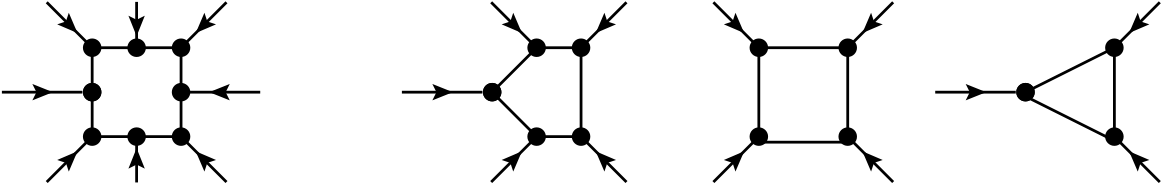}} 
  {\put(-265,28){{ $\in$}}}
   {\put(-250,28){{$\langle$}}}
  {\put(-145,28){{ $,$}}}
    {\put(-85,28){{ $,$}}}
      {\put(-10,28){{$, \,\, \cdots$}}}}%
\end{center}
It is a  theorem of Nickel's \cite{Nickel}, that a general one-loop amplitude can be expressed in terms of a finite and explicit list of `master integrals' of this type.   This topic has generated a
huge amount of work since. However, the corresponding theorem at two and higher-loops  is still unknown. The theory of motivic periods and weights   provides a conjectural answer to this problem  to all loop orders.

A further possible application of the notion of weights is the `leading transcendentality relation between QCD and $N=4$ SYM'  observed in 
 \cite{N4QCD, N4QCD2}. To make sense of this  in general, one of course requires a well-defined notion of weight, and this should be  provided
 by the theory of motivic periods.

\subsection{Application  2: massless $\phi^4$ theory}
Theorem $\ref{thm2}$ states that the Galois conjugate of an amplitude can include  generalised amplitudes of subquotient  graphs, so  
one should expect  arbitrary numerators in the integrand.  Schnetz has made a very bold conjecture which states that the Galois conjugates of graphs in massless $\phi^4$ theory will be  linear combinations of 
\emph{actual} amplitudes of  graphs (i.e., with trivial numerators), and not only that, that these graphs should still have the properties of being  primitive (sub-divergence free) and  in $\phi^4$ theory.

\begin{conj}(O. Schnetz)
The space generated by   the \emph{actual}  motivic amplitudes  $I^{\mm}_G$  of primitive  graphs  $G$ in  massless $\phi^4$ theory (see equation $(\ref{eqn: phi4int}))$ is itself   closed under the action of the cosmic Galois group $\mathcal{C}_{0,0}$. 
\end{conj}

This extraordinary conjecture goes  beyond theorem 2. It has been checked  experimentally  \cite{PanzerSchnetz} for several hundred examples. What this means is the following:

\begin{itemize}
\item Take a primitive graph $G$ for which we know $I_G$ is an MZV or a similar quantity for which we know how to compute the Galois action.
\item Replace the $\zeta$'s with $\zeta^{\mm}$'s  and compute their Galois conjugates.\footnote{Grothendieck's period conjecture for multiple 
zeta values would imply that the operation  of passing from MZV's to motivic MZV's is well-defined. It is completely open.} Check that they are linear combinations of  amplitudes $I_{\gamma}$, defined by $(\ref{eqn: phi4int})$, where 
$\gamma$ is also primitive and of $\phi^4$ type.
\end{itemize}

Consider the following concrete example.
\vspace{0.1in}
\begin{center}
{\includegraphics[width=1.8cm]{K34.png}}%
  \put(-100,22){{ $G=$}}
\end{center}
Recall that its amplitude was
$$I_{G} =32 \times ( \textstyle{ -{216\over 5} \zeta(3,5) -81 \zeta(3)\zeta(5)  + {552\over 5} \zeta(8)})\ ,$$
Now it turns out, since there are few primitive 6-loop graphs with vertices of degree at most four,  that the number 
$\zeta(3,5)$ only ever occurs in the combination
\begin{equation} \label{zeta53combo} 
  \mathbf{-{216\over 5}} \zeta(3,5)   +\mathbf{{552 \over 5} }\zeta(8)\ .
  \end{equation}
This is because of the paucity of graphs compared to the number of MZV's. In fact, there is just one other graph which could conceivably  
give a $\zeta(3,5)$, and it differs from the above by a multiple  of $\zeta(3)\zeta(5)$.  This is the observational input, and is 
a consequence of the small graphs principle (`there are few small graphs').

Now let us  take any amplitude at  higher loop order and
apply the cosmic Galois group. For example, there is a seven-loop graph  \cite{Census} whose amplitude is    
$$I_{P_{7,8}} =  {22383 \over 20 } \zeta(11) + \ldots  + \mathbf{{3024 \over 5 }} \zeta(3)\zeta(3,5) - \mathbf{{7308 \over 5}} \zeta(3) \zeta(8)  $$
with several terms omitted for reasons of space\footnote{Even just a few years ago, computing such an amplitude analytically
would have seemed like science-fiction. Thanks to the  work of Panzer and Schnetz, this is now well within reach.}.
The previous conjecture implies that if the Galois conjugate of the (motivic version) of this amplitude involves
a $\zeta(3,5)$, then it has to occur in the combination $(\ref{zeta53combo})$. 
Indeed, in this case, it successfully predicts the ratio of coefficients
\begin{equation} \label{ratio} 
\mathbf{ {{3024/5} \over {7308/5} }}  = \mathbf{ { 216/5 \over 522/5}}\ ,
\end{equation}
plus three other constraints.  Contrary to what is the usual situation in perturbative quantum field theory, the number of constraints \emph{increases} with the number of loops. There are hundreds of similar  examples, which become increasingly spectacular as the loop order increases.

Now a sceptic might object that one could have guessed the identity $(\ref{ratio})$  empirically simply  by looking at
the coefficients of the amplitude, and do away with all the guff about  motivic periods. 
However, the key point is that there is no unique way to write the amplitude in terms of multiple zeta values because there are many relations between them. Had we written the amplitude using a different choice of MZV's,  then their coefficients would be entirely different, and the relation $(\ref{ratio})$ would be completely obscured.
 Indeed, the basis elements were carefully chosen precisely to make this particular part of the Galois action obvious on the coefficients. 
The deep underlying fact  is that the Galois action is  independent of all possible  choices of basis elements for motivic  multiple zeta values, or put another way, 
the Galois group respects all algebraic relations between motivic multiple zeta values.

\subsection{Some examples of a different nature} I am aware of  a couple of other examples of Galois-theoretic ideas arising in physics. 
The first is in super-string theory, where Stieberger and Schlotterer  \cite{SS} replaced the multiple zeta values which occur in  \emph{tree-level superstring amplitudes} with  motivic multiple zeta values. They went on to  show that this gives an extraordinary simplification of the total amplitude. Their result can be interpreted as a compatibility between tree-level superstring amplitudes and the action of the Galois group of periods.

Lastly,  one can write the known amplitudes for the \emph{anomalous magnetic moment of the electron} in terms of motivic periods. This is beyond the scope of Theorem $\ref{thm2}$ as presently stated because the graphs have apparent infra-red problems, and  we are considering the sum of
all graph amplitudes rather than  individual amplitudes. Nonetheless, the motivic version of ${g-2 \over 2}$ exhibits some  compatibility with the action of the Galois group of periods.  This is equivalent to a number of non-trivial constraints on the known part  of the amplitude (i.e., up to three loops).

Since these examples are quite diverse, it seems reasonable to expect  that a quite  general class of quantum field theories should exhibit some compatibility with the action of the Galois group of periods. Some fundamental questions arise:

 What is the reason for this? Could there  be a physical meaning to the action of the cosmic Galois group\endnotemark[31], or  some part of it?

\subsection{Conclusion}
We saw that   Feynman integrals  and amplitudes form the basis for most predictions in high-energy physics experiments.
They are very far from being understood mathematically, and there are numerous challenges with practical applications. Some problems which 
were not touched upon here are  questions about resummability and existence of renormalisable quantum field theories. 

Grothendieck's deep ideas on motives suggest that there exists a huge symmetry group (motivic Galois group)
acting on period integrals. This is  still highly conjectural, but we can define motivic  periods of graphs unconditionally.

As a result,  we can define a group\endnotemark[32] of hidden symmetries (called the cosmic Galois group\endnotemark[33]) which 
acts on motivic  periods of graphs. It provides an organising principle for the structure of amplitudes and  leads to powerful  constraints, via the stability theorem,  to  all orders in perturbation theory. We are only just beginning to  scratch the surface  of this structure.
\\
\\

\emph{Acknowledgements}. These notes were written whilst the author was a beneficiary of  ERC starting grant 257638.  Many thanks to Erik Panzer for comments and discussions,
and also for the first three graphs in  in the introduction. The figure showing  the 2-loop contributions to $g-2$ were taken from \cite{Hayes}.

\theendnotes

\endnotetext[1]{The sum of Feynman amplitudes is a formal power series in, typically, a coupling constant, and is  called the perturbative expansion.  It is known that this series is divergent, and it is expected that it is  Borel resummable, but this is  not presently known for any renormalisable quantum field theory in four space-time dimensions. }

\endnotetext[2]{The loop number refers to the number of independent cycles, or  equivalently, the rank of the first homology group of the graph.  }

\endnotetext[3]{A fascinating account of the history of the calculation, and the experimental verification, of the anomalous magnetic moment of the electron, of which only   a brief summary is presented here,  is given in \cite{Hayes} and \cite{Styer}.  Feynman's earlier version  of this story  \cite{QED} is also well worth reading.  }

\endnotetext[4]{This is the coefficient of $\alpha^2$   in the perturbative expansion  }

\endnotetext[5]{The higher precision computations for $g-2$ require taking into account QCD effects. It is an extraordinary feature of quantum mechanics that a single experiment,  if left long enough, should in principle generate all the particles of physics.   }
\endnotetext[6]{One expects that one can replace a large number of Feynman integrals with a smaller number of more economical integrals. These  will  be  of a similar type to those considered here, and the theory sketched in this lecture is   likely to apply in these situations also. After all, the perturbative expansion of $2n$-dimensional quantum field theories in flat space-time has coefficients which  are  periods, so it would stand to reason that general methods for understanding period integrals should be of relevance in perturbation theory, no matter by which method they are obtained. Indeed, the completely separate question of what the most efficient integral representation for these periods might  be is very far from being answered at the present time (I  suspect that the rich mathematical structure
 of amplitudes comes from  the fact that they have \emph{several} natural integral representations). What matters here is that there exists \emph{some} integral representation as a period. }

\endnotetext[7]{Euler proved that for $n\geq  1$,   $ \zeta(2n) = -{B_{2n} \over 2}   {(2 i \pi)^{2n} \over (2n)!  } $  where $B_{2n}$ denotes the   $2n^{\mathrm{th}}$ 
Bernoulli number.  }

\endnotetext[8]{From the definitions, one sees that  for $m,n\geq 2$,
$$\sum_{k\geq 1}   {1 \over k^m} \sum_{l\geq 1} {1 \over l^n} = \Big( \sum_{1\leq k<l} + \sum_{1\leq l<k} + \sum_{1\leq k=l} \Big) {1 \over k^m l^n}$$
which implies the relation 
$\zeta(m) \zeta(n) = \zeta(m,n)+ \zeta(n,m) + \zeta(m+n).$ In general, the product of any two multiple zeta values is a linear combination
of multiple zeta values, and the $\Q$-vector space of multiple zeta values forms a ring.
There are many known families of relations between MZV's, but it is still not fully understood how they relate to each other.   }

\endnotetext[9]{One's initial impression on seeing multiple zeta values for the first time is that the definition seems somewhat \emph{ad hoc}. 
 However, it turns out that these numbers are the periods of the most basic possible building blocks which can occur in the cohomology of algebraic varieties called mixed Tate motives. This reason alone  explains why they appear in such a diverse range of subjects.}

 \endnotetext[10]{Passing to Minkowski space involves performing an analytic continuation in external momenta known as a Wick rotation.  There are a  number of technical subtleties, but it is common  practice to  compute amplitudes in  Euclidean space. Note that everything which follows applies to spacetime in any even number of dimensions, not necessarily four.  }

 \endnotetext[11]{This can be the coefficient  of the leading term in $\varepsilon$ in dimensional-regularisation (where one takes $D= 4- 2 \varepsilon$).  This formulation of the Feynman integral allows us, by a sleight of hand (the Gamma pre-factors can be zero) to cover several different cases simultaneously. For 
 the derivation of the parametric Feynman rules with  the correct bells and whistles, see \cite{Angles}. } 

\endnotetext[12]{Every integrand defines a class in the algebraic de Rham cohomology of a suitably-defined  algebraic variety. The latter
is entirely determined by the singular locus of the integrand, which only depends on its denominator. 
The integrals of a basis of cohomology classes could be thought of as the universal periods, or `master integrals' for that graph topology. Changing the numerator, e.g., the type of theory, will change the particular linear combinations of these periods occurring in that theory, but not 
the nature of the periods themselves. The method of `integration by parts' or IBP can be  very profitably reformulated in terms of relative algebraic de Rham cohomology, but this has not yet been done systematically to my knowledge.}

\endnotetext[13]{It is impossible to cite all the recent results in this area, but here follows  a narrow selection
of papers which cover some of the territory. Further references can be found within these papers.
Methods from algebraic geometry: \cite{BEK}, \cite{BlochKreimer},  $p$-adic  and point-counting ideas: \cite{Sta, Stem, BB, K3phi4}. Modular forms occur in \cite{Modularphi4, BroadhurstModular}. For iterated integrals, see  \cite{BrCMP, BrFeyn, Graphical,  Panzer2}, 
for elliptic polylogarithms, \cite{BV, BWA, BSM}. For difference equation and symbolic methods, see \cite{Schneider} and the many references listed there.
  For  surveys on symbols  \cite{Duhr}, differential equations  \cite{Henn}, and computational techniques  \cite{Sm}.
 A sample of geometric approaches: hyperbolic geometry \cite{Davidychev}, cluster algebras \cite{Cluster}, Grassmanian geometry and amplituhedron \cite{Amplituhedron},  twistors  \cite{Twist1,Twist2}, N=4 SYM \cite{Drummond}, spectral curve methods \cite{Wrap}. 
}

\endnotetext[14]{An expression for multiple zeta values as an integral of this form is 
$$\zeta(n_1,\ldots, n_r) =(-1)^r \int_{0\leq t_1 \leq \ldots \leq t_N \leq 1 }  {dt_1 \over t_1- \epsilon_1 } \ldots {dt_N \over t_N - \epsilon_N}$$
where $n_r\geq 2$,  $N= n_1+\ldots + n_r$ is called the weight, and  the sequence of numbers  $(\epsilon_1,\ldots, \epsilon_N)$ is given by $(1, 0^{n_1-1}, \ldots, 1, 0^{n_r-1})$
(a one followed by $n_1-1$ zeros, and so on). This shows that multiple zeta values are periods according to the elementary definition given here. However, to construct motivic versions of multiple zeta values, one needs to express them as periods in the cohomological sense. This 
is considerably more subtle.  
}

\endnotetext[15]{The Galois group $\mathrm{Gal}(\overline{\Q}/\Q)$ is a profinite group: a projective limit of finite groups. It is important to bear in mind that  no element in the Galois group is known  explicitly  except for the identity, and complex conjugation. To know  an element  in $\mathrm{Gal}(\Q)$ is  to 
know how it acts on \emph{all algebraic numbers}, which is probably unreasonable. The group is studied via its  finite-dimensional  representations. The same will be true of the cosmic Galois group: we are more interested in the representations of this group  than the group itself.  }

\endnotetext[16]{One requires the existence of a Tannakian category of mixed motives over $\Q$, and the validity of Grothendieck's period conjecture 
for this category to ensure that the Galois groups are well-defined. Whilst there are various 
approximations to, and   potential candidates for,   the former category, they will not suffice without the full strength of the period conjecture (see footnote 20). The period conjecture  is completely out of reach  except for a handful of  special examples.     }

\endnotetext[17]{It is poor expository style to state a definition  followed by  an example which does not obviously fit the definition. 
However, writing $z= x+iy$, a choice of path  $\gamma$ is given by the polynomial equation $x^2+y^2=1$ and the integrand  is indeed of the required form
$${dz \over z}={(x- iy) (dx + i dy) \over  (x^2+y^2)} \ .$$  
 }

\endnotetext[18]{In the case of an affine scheme $X$ over $\Q$, the algebraic de Rham cohomology is simply given by the cohomology of the 
complex of global regular differential forms  which are defined over $\Q$. }

\endnotetext[19]{See \cite{BrICM} for a definition of motivic periods.  In brief, one fixes a Tannakian category of realisations (at its most simple: the category of triples $(V_B, V_{dR}, c)$ consisting of 
a de Rham  vector space $V_{dR}$ over $\Q$, a Betti vector space $V_B$ over $\Q$, and a comparison isomorphism between their complexifications) and considers the affine ring of isomorphisms from the de Rham to the Betti fiber functor. A motivic period is an element in this ring which comes from geometry: an element $\xi=((V_B, V_{dR}, c), \sigma, \omega)$ where $V_{dR}$, $V_B$ are the de Rham and Betti realisations of the cohomology of a  diagram of algebraic varieties over $\Q$, $c$ the Grothendieck-de Rham comparison isomorphism, and $\omega \in V_{dR}$ and $\sigma \in V_B$.  It defines a function 
$$\phi \mapsto   \sigma(\phi(\omega)) \  : \  \mathrm{Isom}(\omega_{dR}, \omega_B) \To \A^1\ .$$
The theory can be enriched, for example, by requiring $V_{dR}, V_B$ to define a mixed Hodge structure.

In the case of families of periods, one wants a Tannakian category of triples as above where this time $V_{dR}$ is an algebraic vector bundle on some underlying variety $S$ over $\Q$ equipped with an integrable connection with regular singularities, and  $V_B$ a  family of local systems defined over $\Q$  on $S(\C)$. This can be required to define a variation of mixed Hodge structure. The corresponding notion of a family of  motivic periods is a very rich. A family of motivic periods satisfies differential equations and possesses a  monodromy action (continuation along paths), in addition to an action of a Galois group (motivic coaction). }

\endnotetext[20]{There are a number of subtleties when defining the motivic period corresponding to a given period integral. These are often
glossed over.
First of all, the motivic period \emph{a priori} depends on the choice of integral representation (if one assumes all
possible conjectures then it should be independent of  choices). The statement that two motivic periods are equal is a stronger statement 
than the statement that two integrals are equal.  
 Secondly, it is a non-trivial matter to write a period integral given in the elementary sense 
as a pairing between de Rham and Betti cohomology of  algebraic varieties. One can appeal to general (e.g. Hironaka's) theorems on resolution of singularities, but there can be a number of hidden choices. For example, in the relatively simple case of hyperplane configurations 
one wants to express Aomoto integrals as the periods of relative cohomology $H^n(P \backslash A, B \backslash (B\cap A))$
where $A,B \subset P$ are normal crossing divisors in $P$.  Quite aside
from the question of constructing $P$, there can be  some choices involved  in how to define $A$ and $B$. See \cite{Dupont}, where this choice takes the form of a colouring function.
It is not immediately obvious that the motivic periods are independent of all such choices.  
Note also that the choice of compactification $P$ 
is important in determining the Galois conjugates of a motivic period: if $P$ is unnecessarily
complicated, the Galois conjugates that one expects to see will also be unnecessarily complicated. 
For all these reasons, and for practical applications, one really requires a good model for 
a period as a pairing between the cohomology of algebraic varieties which are as simple as possible. The properties of 
those algebraic varieties will determine the nature of the motivic periods.  In  the case of Feynman
amplitudes, this is indeed possible and is the bulk of the work in proving theorem \ref{thm1}.
}

\endnotetext[21]{We have taken the  Galois group to be the Tannaka group with 
respect to the de Rham fiber functor. One could also take the Tannaka group with respect to the Betti functor. In this example, it would give
$ g\log^{\mm}(2) \mapsto  \alpha_g \log^{\mm}(2) +  \beta_g (2  \pi i)^{\mm} \ , $
where $\alpha_g \in \Q^{\times}$ and $\beta_g \in \Q$, and we recognise the term with $2 \pi i$ as the monodromy, or  inherent ambiguity, in the definition of the logarithm function. Thus the number $\log(2)$ `remembers' that it is the value of a multi-valued function, 
and this accounts for part of the action of the Betti Galois group. As the old saying goes, a constant is just a function that  is having a rest.
}

\endnotetext[22]{Using the theory of algebraic groups, we can start to classify motivic periods according to their representation-theoretic properties, inspired by  Galois' proof of  the insolubility of the general quintic by radicals using group theory. Here is a taster.

To any motivic period  $\xi$ we can associate a unique mixed Hodge structure $M(\xi)$.
Thus we can define the \emph{dimension} of a motivic period to be $\dim_{\Q} M(\xi)$, which is also  the dimension of the
representation of $\mathcal{G}$ that $\xi$ generates. Likewise, define the \emph{Hodge polynomial} to be 
the generating series of its Hodge numbers
$$\chi_{\xi}(u,v) = \sum_{p,q}  u^p v^q \dim_{\C} (M^{p,q}(\xi))\ .$$
The \emph{Galois group}  $\mathcal{G}_{\xi}$  of $\xi$ is the largest quotient of $\mathcal{G}$ which acts faithfully on $M(\xi)$.
Finally, define a filtration by  unipotency as follows. The group $\mathcal{G}$ can be written as an extension $$1 \To \mathcal{U} \To \mathcal{G} \To S\To 1$$
where $\mathcal{U}$ is pro-unipotent, and $S$ pro-reductive.  Let $L^i \mathcal{U}$ denote the lower central series of $\mathcal{U}$. I shall say that 
$\xi$ is of \emph{unipotency degree} $\leq k$ if $L^k \mathcal{U}$ acts trivially on  $\xi$. Thus the action of $\mathcal{G}$ on $\xi$ factors through 
$\mathcal{G}/ L^k \mathcal{U}$.  Motivic periods of  unipotency degree $0$ are representations of $S$ and correspond to periods of pure motives. Examples include $2\pi$ or elliptic integrals.  Motivic periods of degree $\leq 1$ correspond to simple extensions.  Examples of these are given by special values of $L$-functions  by Beilinson's conjectures. Here are some examples taken from the text:

$$
\begin{array}{c|c|c|c}
 \hbox{Motivic period } \xi  &  \dim M(\xi) &  \chi_{\xi}(u,v)    & \hbox{degree of unipotency} 
   \\ \hline
  \zetam(2n) \ , \ n\geq 1  &  1  & (uv)^{2n}    &      0 \\
 \log^{\mm}(p)  \ ,  \ \,\, p \hbox{ prime}  & 2    & 1+ uv  &  1  \\
\zeta^{\mm}(2n+1)   \ , \ n\geq 1& 2    & 1+ (uv)^{2n+1}  &  1  \\
 \end{array}
$$
To this one can add the motivic version $\alpha^{\mm}$ of an algebraic number $\alpha \in \overline{\Q} \subset \C$, which has degree of unipotency $0$ and  dimension (and Hodge polynomial equal to) the dimension of the $\Q$-vector space spanned by the Galois conjugates of $\alpha$.
The degree  of $\alpha$ is equal to the number of connected components of   $\Gg_{\alpha^{\mm}}$. 
The above examples are all `mixed Tate', which has the consequence that  $\chi_{\xi}(u,v)$ is a function of $uv$.   There are examples of Feynman integrals
whose motivic periods are not  mixed Tate \cite{K3phi4}.
}

\endnotetext[23]{Understanding the correct role of $\zeta(2)$ has proved to be a sticking point, and is more subtle than one might be tempted to think.
}

\endnotetext[24]{One can deduce that $\zetam(3,5)$ is not a polynomial in any $\zetam(n)$. By Grothendieck's period conjecture,
we expect $\zeta(3,5)$ to be algebraically independent over the $\Q$-algebra generated by the values $\zeta(n)$, $n\geq 2$. 
}

 \endnotetext[25]{It is obvious that generalised Feynman amplitudes are families of periods in the elementary sense given above.
 A version of this result, in the case of  fixed rational masses and momenta but dimensionally-regularised divergent integrals was shown in \cite{BBper} and \cite{BognerWeinzierl}.  To define the motivic periods  one must define the algebraic varieties underlying the Feynman integral in a canonical way,
 and construct  Betti and de Rham (co)homology classes such that the period matches the integral under consideration. This is a much  more precise sense in which Feynman amplitudes are `periods'. For log-divergent graphs in $\phi^4$ with no masses or momentum dependence, this was first done in \cite{BEK}. A more general result with masses and momenta was given in  \cite{IHESyoutube}.}

 \endnotetext[26]{The motivic periods of a graph has a technical definition and may \emph{a priori}  contain 
 more periods than those given by absolutely convergent integrals of the form $(\ref{eqn: GenAmp})$   (one may have to include regularised limits of such integrals).  Hereafter, we use the phrase  `motivic periods' of a graph, or `periods' of a graph  to  denote periods in  this more general sense.}

 \endnotetext[28]{This three-edge calculation enables us to determine $W_2 \HF^{\mm}_{0,0}$, and  implies non-trivial constraints about integrals of the form $(\ref{eqn: phi4int})$ which evaluate to multiple polylogarithms at 2nd, or 6th roots of unity. Examples of such graphs with 7,8, and 9 loops have been found by Panzer and Schnetz \cite{PanzerSchnetz}. For  reasons of exposition, I chose to illustrate the cosmic Galois group in the next section with an example which is simpler, but  depends on a  conjecture. One could replace it with an unconditional, but  more complicated example. }

 \endnotetext[29]{More precisely, there are partial factorisations of the form
 \begin{eqnarray} 
 \Xi_G (q,m )  &= &\Psi_{\gamma} \Xi_{G/\gamma}(q,m) + R^{\Xi, UV}_{\gamma, G}  \nonumber \\
 \Xi_G (q,m )  &= &\Xi_{\gamma}(q,m) \Psi_{G/\gamma}+ R^{\Xi, IR}_{\gamma, G}  \nonumber    
 \end{eqnarray}
 where the first equation holds for all subgraphs $\gamma$, the second for a certain class of subgraph which are mass-momentum spanning 
 (for generic momenta, $\gamma$ is mass-momentum spanning if and only if  $\Xi_{G/\gamma}(q,m)=0$). See \cite{IHESyoutube} for further details.   }

 \endnotetext[31]{One physical source for a  group of symmetries could be the inherent ambiguities in the construction of perturbative quantum theory.
A few candidates jump to mind which may or may  not be related to the cosmic Galois group described here: first of all, there is some (small) freedom in the choice of  renormalisation scheme; 
secondly, the process of Wick rotation involves analytic continuation of multivalued functions and hence there are  possible 
choices of branches to be made; finally, if one believes the theory is Borel resummable but that there are poles on the positive axis in the Borel plane, there can also be choices of the integration path for the Borel transform. This last ambiguity is  related to the theory of resurgence. There are no doubt others. Note that these type of symmetries `as ambiguities' are rather different from the kinds of symmetries `as conserved quantities' of the type occurring  in Noether's theorem.}

 \endnotetext[32]{A certain number of definitions in this talk already appear in the literature with different meanings.
 This is entirely due to the lack of imagination of the author, and the excellent nomenclature already 
 in existence. Certain `motivic amplitudes', for example, are considered in \cite{Cluster} with a very different definition and properties. Likewise,
 Goncharov defined objects called motivic multiple zeta values which are quite different from the objects considered here, and  do not possess a period map to complex numbers, for example. The phrase `cosmic Galois group' was coined by Cartier.  Finally, a different version of the cosmic Galois group was considered in \cite{CM1}. }

\endnotetext[33]{Here is a brief history of ideas surrounding the cosmic Galois group. First of all, based on the experimental 
discovery of multiple zeta values in physics \cite{BK}, Cartier proposed that the Grothendieck-Teichmuller group 
should be related to the renormalisation group in Quantum Field theory \cite{Folle}, \S7, page 42,  and  \cite{CartierInterview}.  This question was taken up in a more precise manner in \cite{Operads}, \S5.2, and by Connes and Marcolli in 
\cite{CM1} and \cite{CM2}, \S5. The first explicit connection between Feynman integrals and motives  was established in \cite{BEK}.
Two missing pieces in this puzzle were provided in \cite{BrMTZ}, where the connection between the motivic Galois group of mixed Tate motives over $\Z$ and multiple zeta values was settled, and, catastrophically,   in \cite{K3phi4}, where it was  shown that in fact not all periods in $\phi^4$ theory were multiple zeta values after all (see also \cite{BB}). Thus the original goal of the cosmic Galois group did not quite turn out as planned, but can  be given a different incarnation using  the concept of motivic periods as described here. In conclusion, the romantic  idea of a cosmic Galois group acting on the constants of physics is likely to be correct; the original guess for what that group was, and how it would act, was too optimistic.}

\bibliographystyle{plain}
\bibliography{main}

\end{document}